\documentclass[twocolumn]{aastex631}

\makeatletter
\newcommand\notsotiny{\@setfontsize\notsotiny{6}{6.5}}
\makeatother

\usepackage{gensymb}
\usepackage{tablefootnote}
\usepackage{float}
\usepackage{graphicx}

\usepackage{indentfirst}
\usepackage{titletoc}
\usepackage{amssymb}
\usepackage{placeins}
\usepackage{stackengine}
\usepackage{pifont}

\newcommand{\xmark}{\ding{53}}%
\usepackage{color}

\begin{document}
\title{A Comprehensive Investigation of Gamma-Ray Burst Afterglows Detected by \textit{TESS}}

\correspondingauthor{Hugh Roxburgh}
\email{hro52@uclive.ac.nz}

\author[0009-0001-6992-0898]{Hugh Roxburgh}
\affiliation{School of Physical and Chemical Sciences — Te Kura Matū, University of Canterbury, Private Bag 4800, Christchurch 8140, \\ Aotearoa, New Zealand}

\author[0000-0003-1724-2885]{Ryan Ridden-Harper}
\affiliation{School of Physical and Chemical Sciences — Te Kura Matū, University of Canterbury, Private Bag 4800, Christchurch 8140, \\ Aotearoa, New Zealand}

\author[0009-0003-8380-4003]{Zachary G. Lane}
\affiliation{School of Physical and Chemical Sciences — Te Kura Matū, University of Canterbury, Private Bag 4800, Christchurch 8140, \\ Aotearoa, New Zealand}

\author[0000-0002-4410-5387]{Armin Rest}
\affiliation{Department of Physics and Astronomy, The Johns Hopkins University, Baltimore, MD 21218, USA}
\affiliation{Space Telescope Science Institute, 3700 San Martin Drive, Baltimore, MD 21218, USA}

\author[0009-0005-0556-3886]{Lancia Hubley}
\affiliation{School of Physical and Chemical Sciences — Te Kura Matū, University of Canterbury, Private Bag 4800, Christchurch 8140, \\ Aotearoa, New Zealand}

\author[0000-0002-0476-4206]{Rebekah Hounsell}
\affiliation{The University of Maryland Baltimore County, 1000 Hilltop Cir, Baltimore, MD 21250, USA}
\affiliation{NASA Goddard Space Flight Center, 8800 Greenbelt Road
Greenbelt, MD 20771, USA}

\author[0000-0001-5233-6989]{Qinan Wang}
\affiliation{Bloomberg Center for Physics and Astronomy, Johns Hopkins University, Baltimore, MD 21218, USA}

\author[0000-0001-6395-6702]{Sebastian Gomez}
\affiliation{Space Telescope Science Institute, 3700 San Martin Drive, Baltimore, MD 21218, USA}

\author[0000-0002-2361-7201]{Justin Pierel}
\affiliation{Space Telescope Science Institute, 3700 San Martin Drive, Baltimore, MD 21218, USA}

\author[0000-0002-0786-7307]{Muryel Guolo}
\affiliation{Bloomberg Center for Physics and Astronomy, Johns Hopkins University, Baltimore, MD 21218, USA}

\author[0000-0002-3825-0553]{Sofia Rest} \affiliation{Bloomberg Center for Physics and Astronomy, Johns Hopkins University, Baltimore, MD 21218, USA}

\author[0000-0002-8436-5431]{Sophie von Coelln} 
\affiliation{Bloomberg Center for Physics and Astronomy, Johns Hopkins University, Baltimore, MD 21218, USA}


\begin{abstract}


Gamma-ray bursts produce afterglows that can be observed across the electromagnetic spectrum and can provide insight into the nature of their progenitors. While most telescopes that observe afterglows are designed to rapidly react to trigger information, the \textit{Transiting Exoplanet Survey Satellite (TESS)} continuously monitors sections of the sky at cadences between 30 minutes and 200 seconds. This provides \textit{TESS} with the capability of serendipitously observing the optical afterglow of GRBs. We conduct the first extensive search for afterglows of known GRBs in archival \textit{TESS} data reduced with the \texttt{TESSreduce} package, and detect 11 candidate signals that are temporally coincident with reported burst times. We classify 3 of these as high-likelihood GRB afterglows previously unknown to have been detected by \textit{TESS}, one of which has no other afterglow detection reported on the Gamma-ray Coordinates Network. We classify 5 candidates as tentative and the remainder as unlikely. Using the \texttt{afterglowpy} package, we model each of the candidate light curves with a Gaussian and a top hat model to estimate burst parameters; we find that a mean time delay of $740\pm690\,$s between the explosion and afterglow onset is required to perform these fits. The high cadence and large field of view make \textit{TESS} a powerful instrument for localising GRBs, with the potential to observe afterglows in cases when no other backup photometry is possible, and at timescales previously unreachable by optical telescopes.

\end{abstract}



\section{Introduction} \label{sec:intro}

Gamma-ray bursts (GRBs) are the most powerful explosions we observe in the universe, surpassing all other catastrophic events in terms of electromagnetic luminosity \citep{Gil22}. Peaking at $\gamma$-ray wavelengths, these bursts release rapid jets of radiation that can carry more energy in the span of a few seconds than the Sun will in its entire 10 billion year lifetime \citep{Fis97}.

\begin{figure}
    \centering
    \includegraphics[width=0.5\textwidth]{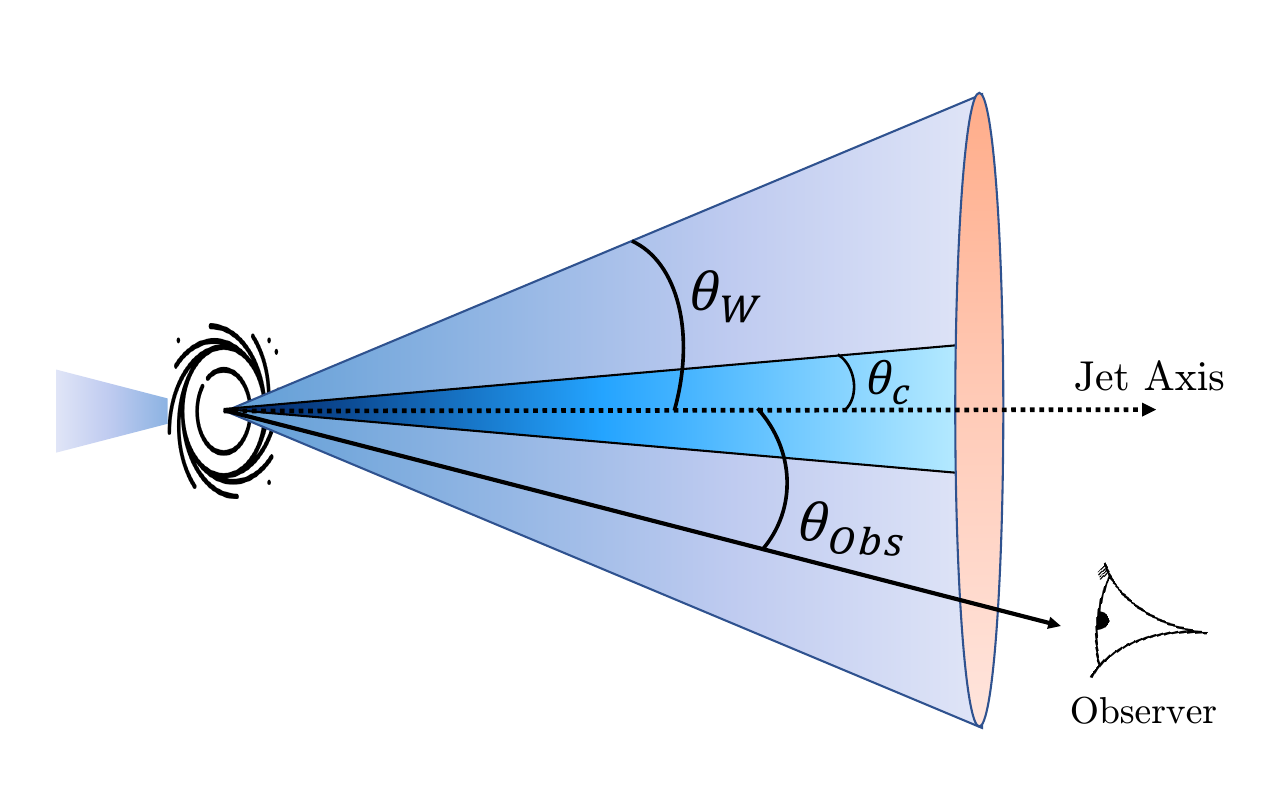}
    \caption{Simplified schematic of a GRB, displaying three key geometric parameters. $\theta_{\rm w}$ is the outer wing truncation angle, $\theta_{\rm obs}$ is the angle from the jet axis to the observer, and $\theta_{\rm c}$ is the angle of the central jet core.}
    \label{fig:GRB_Schematic}
\end{figure}

GRBs emit most of their radiation out in narrow conical-shaped jets \citep{Sar99} which can be described using a number of geometric parameters, most notably $\theta_{\rm obs}$, $\theta_{\rm c}$, and $\theta_{\rm w}$, as shown in Fig.~\ref{fig:GRB_Schematic}. These represent the angles measured from the burst axis to the observer, jet core, and outermost jet boundary respectively. The majority of the energy released by GRBs is concentrated within the region defined by the jet core, $\theta_{\rm c}$, whereas beyond the boundary, $\theta_{\rm w}$, the energy emission is virtually negligible. Consequently, in general $\theta_{\rm obs} < \theta_{\rm w}$.

Beyond their geometric properties, GRBs are also described by their durations and energies. The duration of a GRB is quantified using the parameter $T_{90}$, which is defined as the time interval during which the cumulative number of detected photons increases from 5$\%$ to 95$\%$ of the total recorded counts \citep{Kou93}. Based on this definition, GRBs are classed into two subcategories: short ($T_{90} \le 2\,$s) and long ($T_{90} > 2\,$s). The former have generally been attributed to merger events between extremely compact celestial bodies \citep{Eic89}, such as neutron stars or black holes. There has been significant evidence over the past decade to support this hypothesis - most notably the detection of a short GRB accompanying the gravitational wave event GW170817 \citep{Abb17} and kilonova AT 2017gfo.

Long GRBs constitute approximately 70$\%$ of all GRB detections and have durations that can range up to several minutes, with an average $T_{90}$ of $\sim 20\,$s \citep{Kou93}. Because these events have a strong correlation with galaxies exhibiting high rates of star formation, they are strongly associated with hypernovae explosions, resulting from the core collapse of supermassive stars \citep{Woo93}. However, recent studies have revealed a more significant overlap in progenitor systems than previously expected, with a supernova accompanying the short GRB200826A \citep{Zhang_2021, Ahumada_2021, Rossi22} and a kilonova accompanying the long GRB211211A \citep{Ras22, Tro22}. 

The prompt GRB emission is followed by an afterglow at longer wavelengths that persists on timescales ranging from minutes to months in rare cases \citep{Mes97}. These afterglows are generated during the interaction of the relativistic outflow and the circumburst medium (CBM), where external shocks decelerate the charged particles and release synchrotron radiation \citep{Piran}. Due to their dependency on the CBM and inhomogeneous magnetic fields around the source, we can use detailed observations of afterglow light curves to deduce properties of the progenitor and surrounding region \citep{Wan18}. Open-source packages such as \texttt{afterglowpy} can be used to model these properties \citep{Rya20}.

The rapid nature of GRB emission makes observing them challenging. GRBs are initially detected by `trigger' instruments, primarily the Gamma-ray Burst Monitor on board the \textit{Fermi Gamma-ray Space Telescope} \citep[\textit{Fermi};][]{Mee08} and the Burst Alert Telescope on board the \textit{Neil Gehrels Swift Observatory} \citep[\textit{Swift};][]{Bar05}. These instruments attempt to localise the burst during its prompt emission, before communicating information through the Gamma-ray Coordinates Network (GCN) for follow up imaging by ground based telescopes. In cases where a burst is poorly localised, largely a result of \textit{Fermi's} poor initial estimate and lack of rapid follow up X-ray detector, the GCN is unable to mobilise adequately. This may prevent broadband observations from being made \citep{Ber19}. 

Telescopes with high-cadence observations over large areas can provide unique data for studying GRBs through serendipitous observations. The \textit{Transiting Exoplanet Survey Satellite} \citep[\textit{TESS}, ][]{Ric15} is one such telescope, with a $2160\rm \,deg^2$ field of view (FoV) and near continuous observation over each $\sim$27 day `sector'. With cadences decreasing from 30 minutes (2018-2020), to 10 minutes (2020-2022), to 200 seconds (2022-present), \textit{TESS} is becoming an increasingly ideal instrument for serendipitous detection of GRB afterglows, particularly with respect to obtaining observations during the rising phase. The capability of \textit{TESS} to capture these afterglows was shown in the analysis of GRB191016A \citep{Smi21}. While well-localised GRBs such as GRB191016A could be readily identified and studied, the large and challenging data volume have limited the extent to which \textit{TESS} observations could be utilised for GRB analysis.

In this paper we build on the \texttt{TESSreduce} pipeline \citep{Rid21} to conduct the first systematic search for afterglows originating from known GRBs that were serendipitously observed by \textit{TESS}. We also model resulting light curves using \texttt{afterglowpy}. Section~\ref{sec:data} describes the \textit{TESS} data and the source catalogues we utilise in this project. Section~\ref{sec:analysis} describes the pipeline we implement to search for afterglows, and presents the results. Section~\ref{sec:discussion} discusses the likelihood that the candidates are GRB afterglows. We also discuss how these results support the use of \textit{TESS} for detecting and characterising rapid extragalactic transients.

\section{Data} \label{sec:data}
\subsection{Gamma-Ray Burst Coordinates Network}

We use the GRBweb event list\footnote{\url{https://user-web.icecube.wisc.edu/~grbweb_public/index.html}} compiled from sources in the GCN \citep{Kienlin2020, Lien2016, Ajello2019, Hurley2013, Barthelmy2000} as a catalogue of all known bursts over the lifetime of \textit{TESS} operations. With the detection time, coordinate, and associated 1-dimensional position error (1$\sigma$), we check for \textit{TESS} observations that overlap in time and with the 2$\sigma$ error region. Due to \textit{TESS}'s wide FoV, we have the capacity to search for afterglows with large positional errors that were otherwise not observed. 

From the total list of 1444 GRBs that have occurred since the launch of \textit{TESS} (as of February 2023), we find that 69 have coincidental observations. Of these 69 bursts, only GRB191016A was known to have been detected by \textit{TESS} prior to this study. One other GRB, occurring in March 2023, has subsequently been detected and analysed \citep{Fau23}. Among our sample, 56 are long GRBs with $T_{90}>2$s, and 4 are short GRBs with $T_{90}<2$s (GRB180727A, GRB190507C, GRB221004A, GRB221120A); the remaining 9 GRBs have no solid constraint on $T_{90}$. 

\subsection{\textit{TESS}}

\textit{TESS} observes $24\degree \times 96\degree$ sections of the sky continuously for $\sim$27 days in a highly elongated orbit around Earth. During its lifetime, \textit{TESS} has recorded Full-Frame Images (FFIs) at three cadences: 30~minutes (25\textsuperscript{th} July 2018 -- July 4\textsuperscript{th} 2020),  10~minutes (July 5\textsuperscript{th} 2020 -- 31\textsuperscript{st} August  2022), and  200~seconds (31\textsuperscript{st} August  2022 -- present). It is fitted with a very broad bandpass ranging from the optical to near-IR ($\rm \sim600-1050\,$nm), and has a low angular resolution of 21 arcseconds per pixel.

Each $\sim$27 day observation period constitutes a \textit{TESS} `sector'. The time series of calibrated \textit{TESS} FFIs for each sector are made available through the MAST archive\dataset[doi:10.17909/0cp4-2j79]{http://dx.doi.org/10.17909/0cp4-2j79}. The images are released as $2078 \times 2136$ pixel files, though only $2048 \times 2048$ contain captured data. We process all \textit{TESS} data with \texttt{TESSreduce} to remove the scattered light background that affects cameras near the ecliptic, as described in section 3 of \citet{Rid21}.
\newpage
\section{Analysis} \label{sec:analysis}

\subsection{Pipeline}\label{subsec:pipe}

For rapid transients like GRB afterglows, the high-cadence \textit{TESS} data can provide multiple data points during the event light curve. In this search we examine all available \textit{TESS} data that overlaps with the reported GRB $2\sigma$ positional error at the time of the burst to identify any potential counterparts. Using the following criteria we search all available \textit{TESS} data for GRB afterglows. For a pixel to be considered as containing a candidate GRB afterglow it must meet the following conditions:
\begin{enumerate}
    \item have a local maximum within 1~hour of the burst trigger, as we expect the brightness to decay exponentially following the rapid rise;
    \item exceed a brightness threshold of median$+4\sigma$, set by the median and standard deviation of the pixel's light curve 12 hours before and after the trigger;
    \item remain above the brightness threshold for 2 consecutive exposures.
\end{enumerate}

The localisations of the GRBs vary from arcminutes to degrees. In all cases we determine which \textit{TESS} cameras and CCDs provide coverage of the GRBs and download the corresponding time series of FFIs. We create data cubes, or Target Pixel Files (TPFs), from the FFIs using the \texttt{astrocut} package for \texttt{python} \citep{Bra19}.

\begin{figure*}[t]
    \centering
    \hspace{-1cm}

    \gridline{\fig{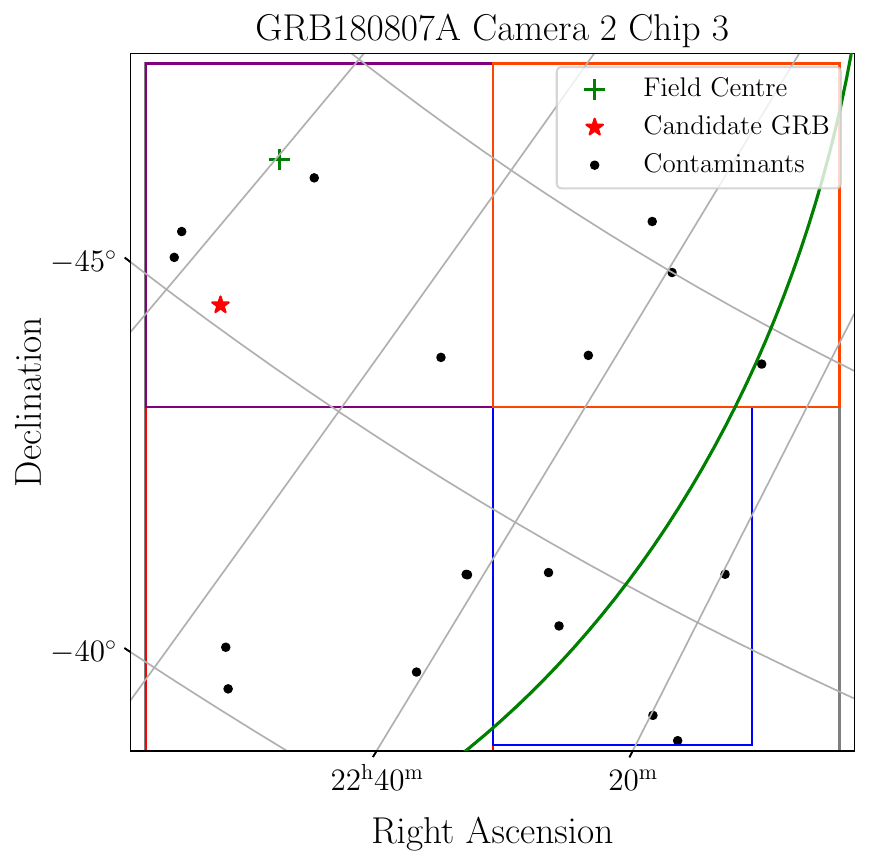}{0.55\textwidth}{}
          \fig{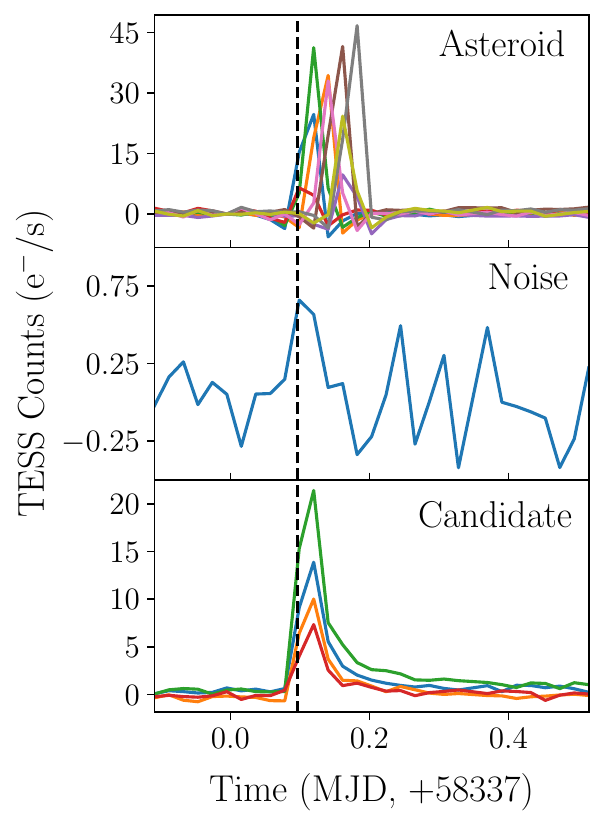}{0.385\textwidth}{}}

    \vspace{-0.5cm}
    \caption{Detection pipeline as performed on GRB180807A. (Left) Events detected by the pipeline on camera 2 chip 3 coincident with the time of GRB180807A. The green curve displays the boundary of the 2$\sigma$ error distance from the estimated location of the GRB as reported by the GCN, which is represented by the green plus in the top left. The coloured boxes display the regions of the cutouts we generate and operate upon individually. Each black point represents an `event' detected by the pipeline; these are the signals of asteroids (eg. top right, each light curve corresponds to a single pixel) or random noise (eg. middle right). The red star displays the location of the candidate GRB signal (bottom right). Note that the dashed line in each of the light curves on the right represents the reported detection time of GRB180807A.}
    \label{fig:cands}
\end{figure*}

After constructing the full frame TPFs, we cut them according to their error regions using \texttt{astrocut}. Due to computational limitations of \texttt{TESSreduce}, TPFs larger than $1.6\times 10^6$~pixels must be segmented into smaller TPFs. These segments can have a maximum size of $1300\times1300$~pixels and share a buffer zone of 20~pixels. An example of a segmented region can be seen in the left panel of Fig.~\ref{fig:cands}.

In all cases we perform photometric reduction on the cutouts using \texttt{TESSreduce}. Due to the size of the files, the more rigorous steps normally applied by \texttt{TESSreduce} are omitted, such as the secondary background correlation correction method and flux calibration. These processes are not required for the search procedures we implement, but are used when extracting the final light curve once a candidate is identified.

Using the criteria outlined above for event detection, we can search for transients in the reduced data. Pixels that meet the detection criteria are flagged as candidates. Neighbouring pixels that trigger at the same time are gathered and considered the same event. Comparing the light curves from each pixel in an event provides a useful diagnostic for determining the validity of the event. 

All events in a GRB field are manually vetted for legitimacy. An example field for GRB180807A with identified candidates is shown in the left panel of Fig~\ref{fig:cands}. The majority of these detections are easily identifiable as asteroids, characterised by neighbouring pixels peaking in brightness at different times and thus signifying a source moving across the frame (Fig~\ref{fig:cands} top right), or as peculiarities in the data reduction due to their light curve shape and number of triggered pixels (Fig~\ref{fig:cands} middle right). To pass the visual vetting stage, candidates must have a light curve shape consistent with a GRB afterglow -- a rapid rise followed by a power law decay. Objects that pass visual vetting are reduced with the full \texttt{TESSreduce} pipeline, which reduces a $90\times90$~pix image cutout using TESScut. We then localise the source with the \texttt{photutils} \texttt{centroid\_sources} function on the frame with peak candidate flux  \citep{Bradley2022}. 

As a final vetting, stage we check deep imaging of the regions with the Pan-STARRS\footnote{\url{http://ps1images.stsci.edu/cgi-bin/ps1cutouts}} and SkyMapper DR2\footnote{\url{https://skymapper.anu.edu.au/image-cutout/}} image cutout services for the northern and southern targets respectively. Candidates whose coordinates appear near to visible sources are included in our findings, though we note their limited reliability. In general, we only find at most one candidate per GRB that satisfies our selection criteria for follow-up vetting.

\subsection{\textit{TESS} Candidate GRB Afterglows}

Table~\ref{tab:detections} presents a list of candidate afterglow signals that were temporally coincident with known GRBs reported on the GCN, alongside the known afterglow of GRB191016A. The light curves of these candidates are displayed in Figs.~\ref{fig:lcs} and \ref{fig:lcs_bad}, and are classed into three groups based on our confidence in their legitimacy. Our 

\twocolumngrid  
\clearpage      

\onecolumngrid

\begin{deluxetable}{cccccccccc}[t!]
    \tabletypesize{\scriptsize}
    \tablecaption{GRB optical afterglow candidates detected by \textit{TESS}. \label{tab:detections}}
    \tablehead{
    \colhead{GRB Name} & \colhead{Detection Time} & \multicolumn{2}{c}{Estimated Position} & \colhead{Position Error} & \multicolumn{2}{c}{\textit{TESS} Coverage} & \colhead{} & \multicolumn{2}{c}{Candidate Position}\\
    \cline{3-4}
    \cline{6-7}
    \cline{9-10}
    \colhead{} & \colhead{(MJD)} & \multicolumn{1}{r}{R.A. (\degree)} & \multicolumn{1}{c}{Dec. (\degree)} & \colhead{($1\sigma$, \degree)} & \colhead{Sector} & \colhead{[Cam,CCD]$^\ast$} & \colhead{} & \colhead{R.A. (\degree)} & \colhead{Dec. (\degree)}
    } 
    \startdata
    GRB180807A & 58337.096966 & 349.93 & -47.84 & 6.643 & 1 & \textbf{[2,3]},[1,2],[2,2] & & $348.786\pm0.001$ & $-45.328\pm0.001$\\ 
    GRB181208A & 58460.187968 & 104.11 & -66.66 &  3.553 & 5 & \textbf{[4,4]},[4,1] & & $113.363\pm0.005$ & $-70.090\pm0.002$\\
    GRB190117B & 58500.368548 & 105.27 & -20.57 & 4.224 & 7 & \textbf{[2,1]},[2,2],[2,4] & & $100.0913\pm0.002$ & $-22.231\pm0.002$\\ 
    GRB190308A & 58550.923458 & 171.94 & -21.73 & 7.134 & 9 & \textbf{[1,1]},[1,2],[1,3],[1,4] & & $162.091\pm0.001$ & $-16.150\pm0.002$\\ 
    GRB190723A & 58687.308530 & 289.47 & 25.22 & 7.633 & 14 & \textbf{[1,2]},[1,1],[1,3],[1,4] & &$284.013\pm0.002$ & $33.983\pm0.002$ \\ 
    GRB191016A$^\dag$ & 58772.172917 & 30.2695 & 24.5099 & $\sim0$ & 17 & \textbf{[1,4]} & & - & - \\
    GRB200111A & 58859.632731 & 107.87 & 32.55 & 4.300 & 20 & \textbf{[1,3]},[1,4] & &$99.293\pm0.002$ & $37.079\pm0.002$\\ 
    GRB200412B & 58951.381019 & 277.48 & 61.78 & 0.500 & 23 & \textbf{[4,3]} & &$278.313\pm0.003$ & $62.532\pm0.001$\\
    GRB210114A & 59228.894970 & 115.81 & -23.40 & 7.972 & 34 & \textbf{[2,4]},[2,1],[2,2],[2,3] & &$118.792\pm0.002$ & $-12.431\pm0.002$\\
    GRB210317A & 59290.380891 & 154.22 & -64.94 & 2.784 & 36 & \textbf{[3,4]},[3,1] & &$149.747\pm0.005$ & $-64.10\pm0.002$\\
    GRB220310A & 59648.019410 & 168.61 & 23.45 & 0.183 & 49 & \textbf{[1,3]} & &$168.260\pm0.001$  & $23.252\pm0.001$\\
    GRB220514B & 59713.975243 & 245.79 & 61.65 & 1.000 & 51 & \textbf{[4,1]}& &$246.550\pm0.003$ & $61.042\pm0.001$\\ 
    \enddata
    \centering
    \tablenotetext{\dag}{Previously known detections prior to this study.} \tablenotetext{\ast}{Bold indicates camera, CCD combination containing candidate. }
\end{deluxetable}

\begin{figure}[t!]
    \centering
    \includegraphics[width=\textwidth]{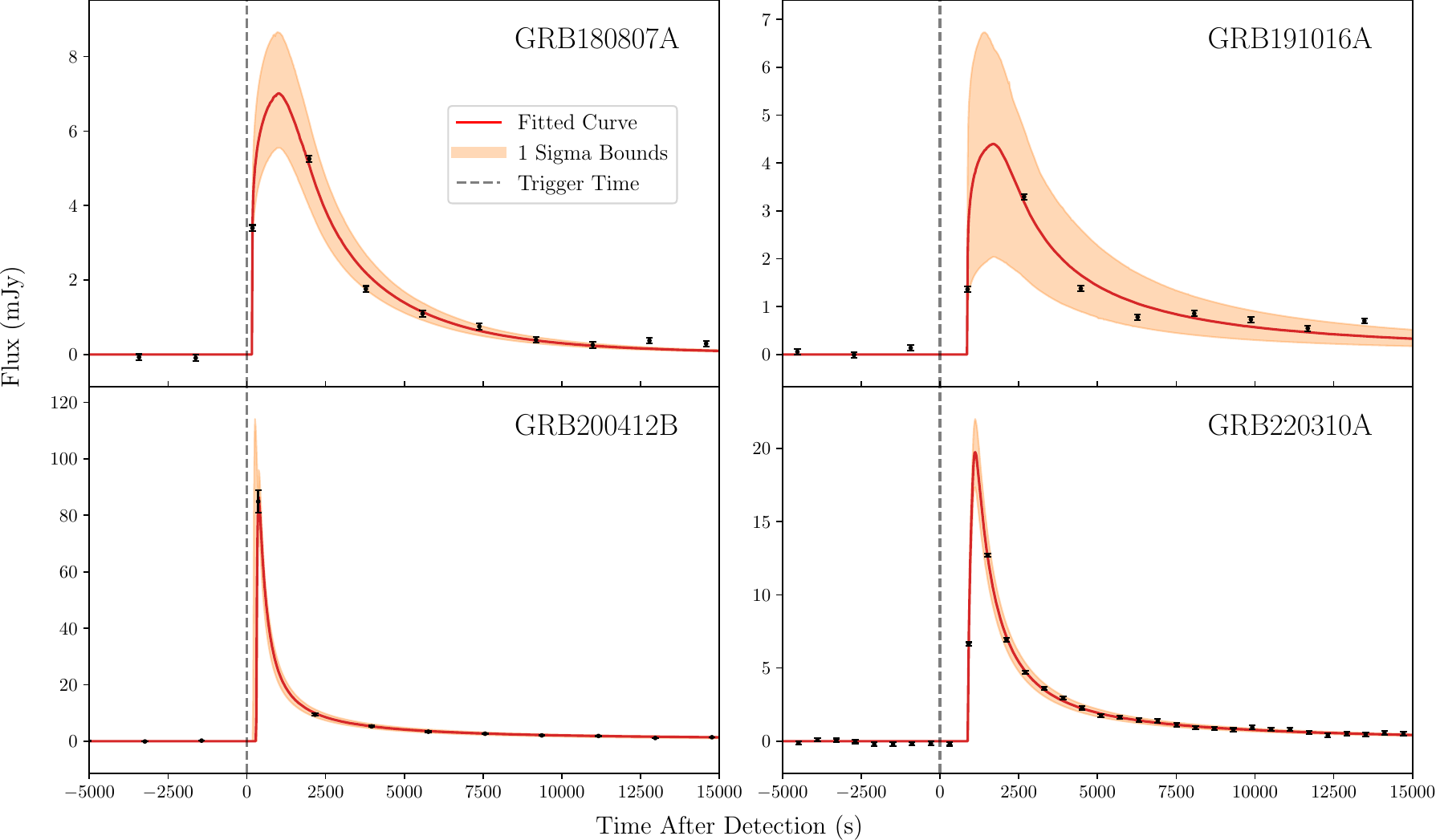}
    \caption{Light curves and overlaid models of optical afterglows observed by \textit{TESS}. GRB191016A (top right) is a confirmed detection; the remaining are high-likelihood candidates discovered by our pipeline. Each red line represents the best fit top hat model generated with \texttt{afterglowpy}, with the orange shaded regions representing the $1\sigma$ limits; see Section~\ref{subsec:model} for details. The time axis is presented with respect to the detection time (dashed line) of each GRB as reported by the GCN. \label{fig:lcs}}
\end{figure}

\begin{figure}[t!]
    \centering
    \includegraphics[width=\textwidth]{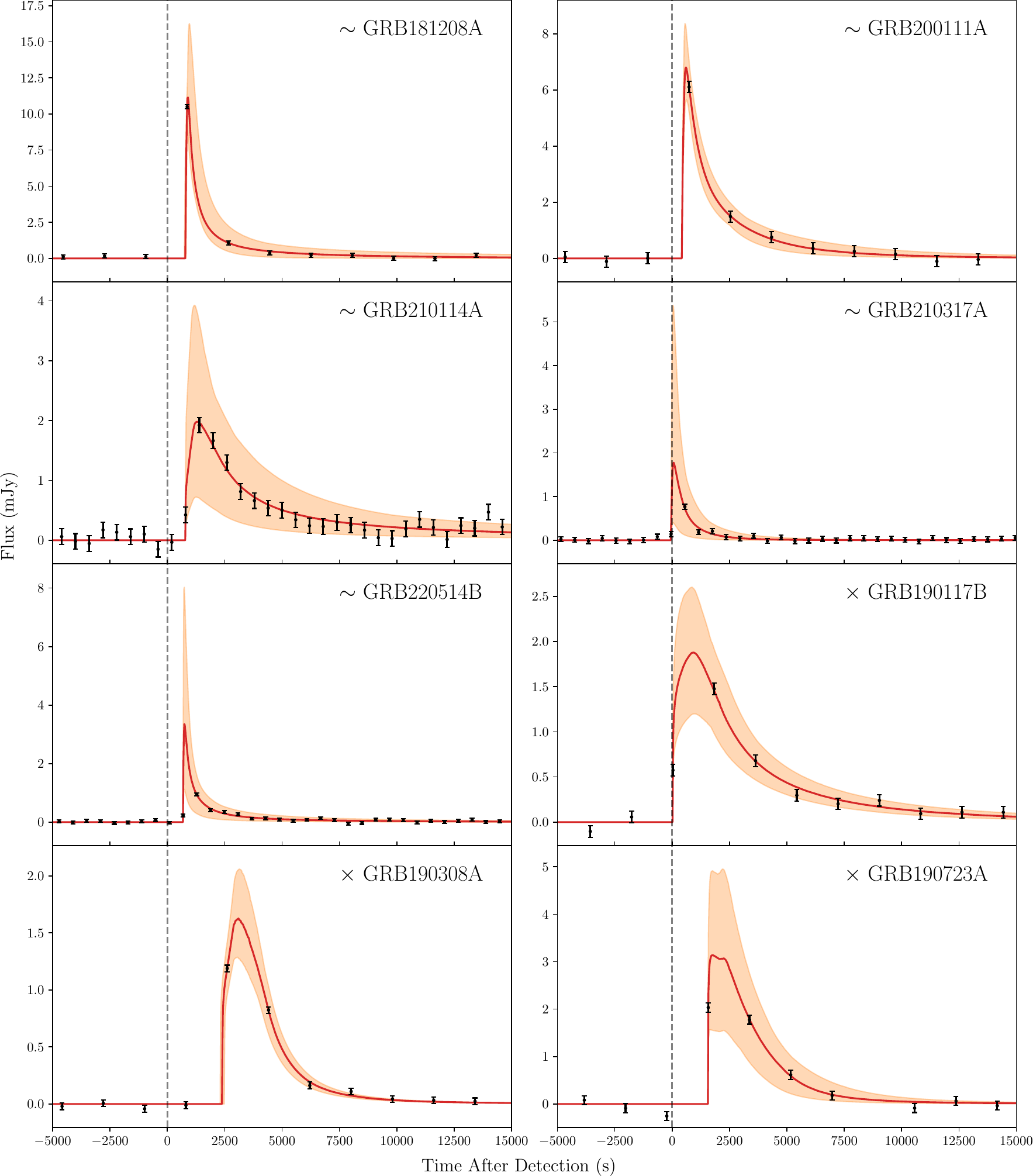}
    \caption{Continued candidate light curves for detections we consider less likely to originate from their respective GRBs. Light curves with $\sim$ preceding their names are tentative candidates, whereas those with $\times$ preceding their names are our least likely candidates; see Section~\ref{subsec:confidence} for details.\label{fig:lcs_bad}}
\end{figure}

\clearpage
\twocolumngrid


\noindent reasoning for these classifications is discussed in Section~\ref{subsec:confidence}. The light curves are extracted with the complete \texttt{TESSreduce} reduction process, where we use a custom aperture for each GRB that maximises the collected flux while avoiding contamination from nearby sources due to the large pixel size. 

\subsection{Modelling with \texttt{afterglowpy}} \label{subsec:model}

Also displayed in Fig.~\ref{fig:lcs} are models generated with the \texttt{afterglowpy} package for \texttt{python}. This package calculates semi-analytic afterglow light curves for input wavelengths and times since explosion. These light curves are produced by modelling synchrotron emission caused by the forward shock, incorporating the jet geometry and viewing angle. The models are dependent on a range of physical parameters: the isotropic-equivalent energy ($E_{\rm iso}$ in ergs), the redshift in the cosmic microwave background frame ($z_{\rm cmb}$), the jet angles for the observer, core, and wing ($\theta_{\rm obs}$, $\theta_{\rm c}$, $\theta_{\rm w}$, all in radians), and four variables describing the density and energy distribution of the simulated circumburst medium (see \cite{Rya20} for details). All of our models assume a spatially-flat $\Lambda$CDM cosmology with $\rm H_{0}=70.0\,$kms$^{-1}$Mpc$^{-1}$ and $\rm \Omega_{m_0}=0.3$. Finally, we include an additional parameter $t_0$, which is the time in seconds from the GRB trigger to the afterglow onset.

It should be noted that while \texttt{afterglowpy} provides an easy way to model GRB afterglows, it has some shortcomings. Our additional $t_0$ parameter is required to account for its lack of initial coasting phase, and it also does not incorporate reverse shocks, which may make early light curves unreliable. Furthermore, it assumes a constant density for the CBM, and therefore is unable to accurately model afterglows that originate in environments with wind, as expected in collapsars \citep{Chevalier_2000,Granot_2002}. 


With only the \textit{TESS} data, the models suffer from high levels of degeneracy between parameters. This degeneracy is particularly significant between the $E_{\rm iso}$ and $z_{\rm cmb}$ parameters, largely due to a lack of redshift constraint resulting from the absence of spectroscopic or host galaxy observations. While one could impose constraints on $z_{\rm cmb}$ by assuming a lower bound based on the magnitude limits of PS1 and SkyMapper, we chose not to do so due to the wide range of plausible host magnitudes when considering faint dwarf galaxies \citep{Simon_2019}. Consequently, broad and uniform distributions were selected as parameter priors; these distributions are displayed in the third row of Table \ref{tab:model_params}. 

Because of this degeneracy, both the \texttt{emcee} \citep{emcee13} and \texttt{nestle} \citep{skilling04, Mukherjee06, shaw07, feroz09} packages failed to converge for the full parameter fit with a Gaussian jet geometry. Likewise, a simplified model, setting $n_0=1$, $p=2.2$, $\varepsilon_{\rm E}=0.1$, and $\varepsilon_{\rm B}=0.01$ within a top hat jet geometry, failed to converge. As an alternative, we used \texttt{scipy}'s \texttt{curve\_fit} (CF) algorithm \citep{2020SciPy-NMeth} alongside a non-linear least squares optimisation. In this process, we first fit the parameters through CF and then used the output as the initial parameters for the least squares optimisation. Through trial fitting, we found that the simplified top hat model provided results that exhibited negligible differences to the full parameter Gaussian jet model whilst requiring significantly lower computational overhead, so we use the simplified model for all fits. The light curves for these fits are shown in Figs.~\ref{fig:lcs} and \ref{fig:lcs_bad}, with the corresponding parameters shown in Table~\ref{tab:model_params}. From this model fitting, it is clear that additional observations at different wavelengths are required to break the degeneracies we encounter. 


\begin{deluxetable*}{ccccccc}[t]
    \centering
    \tabletypesize{\scriptsize}
    \tablecaption{The parameters used for modelling each GRB with a tophat model in \texttt{afterglowpy}. Four parameters are held constant, namely $n_0 = 1$, $p = 2.2$, $\varepsilon_{\rm E} = 0.1$, and $\varepsilon_{\rm B} = 0.01$. \label{tab:model_params}}
    
    \tablehead{
    \colhead{GRB Name} & \colhead{Cadence} & \colhead{$log_{10}(E_{\rm iso})$} & \colhead{$z_{\rm cmb}$} & \colhead{$\theta_{\rm c}$} & \colhead{$\theta_{\rm obs}$} & \colhead{$t_0$} \\
    \cline{3-7}
    \colhead{} & \colhead{(Minutes)} & \colhead{(erg)} & \colhead{} & \colhead{(deg)} & \colhead{(deg)} & \colhead{(s)} \\
    \cline{3-7}
    \colhead{} & \colhead{} & \colhead{[45,57]} & \colhead{[0,4]} & \colhead{[0,90]} & \colhead{[0,90]} & \colhead{[-2000,3000]}
    } 
    
    \startdata
    GRB180807A & 30 & 53.46 $\pm$ 0.05 & 1.79 $\pm$ 0.14 & 1.03 $\pm$ 0.06 & 0.458 $\pm$ 0.06 & 170.2 $\pm$ 0.2  \\ 
    GRB181208A & 30 & 50.78 $\pm$ 0.96 & 0.06 $\pm$ 0.03 & 3.38 $\pm$ 4.01 & 2.69 $\pm$ 1.72 & 789.9 $\pm$ 41.0 \\ 
    GRB190117B & 30 & 53.28 $\pm$ 0.15 & 3.02 $\pm$ 0.87 & 1.15 $\pm$ 0.06 & 0.06 $\pm$ 0.06 & 32.9 $\pm$ 0.1 \\
    GRB190308A & 30 & 52.79 $\pm$ 0.06 & 1.58 $\pm$ 0.24 & 0.92 $\pm$ 0.06 & 0.00 $\pm$ 0.06 & 2399.1 $\pm$ 98.7\\
    GRB190723A & 30 & 53.25 $\pm$ 0.19 & 1.71 $\pm$ 0.45 & 0.69 $\pm$ 0.06 & 0.46 $\pm$ 0.06 & 1565.2 $\pm$ 5.6 \\
    GRB191016A & 30 & 53.45 $\pm$ 0.26 & 2.29 $\pm$ 1.22 & 12.20 $\pm$ 0.06 & 11.75 $\pm$ 0.40 & 874.6 $\pm$ 0.6 \\
    GRB200111A & 30 & 51.25 $\pm$ 0.33 & 0.13 $\pm$ 0.06 & 2.98 $\pm$ 0.69 & 0.00 $\pm$ 0.17 & 431.4 $\pm$ 1.4 \\
    GRB200412B & 30 & 50.20 $\pm$ 0.38 & 0.01 $\pm$ 0.01 & 50.82 $\pm$ 7.45 & 40.22 $\pm$ 0.06 & 299.7 $\pm$ 104.5 \\ 
    GRB210114A & 10 & 52.47 $\pm$ 0.43 & 0.96 $\pm$ 0.49 & 6.02 $\pm$ 0.46 & 5.04 $\pm$ 0.57 & 794.3 $\pm$ 0.7 \\ 
    GRB210317A & 10 & 50.66 $\pm$ 0.30 & 0.13 $\pm$ 0.07 & 2.12 $\pm$ 0.06 & 0.63 $\pm$ 0.69 & -46.1 $\pm$ 0.1 \\ 
    GRB220310A & 10 & 51.82 $\pm$ 0.06 & 0.14 $\pm$ 0.01 & 3.04 $\pm$ 0.11 & 2.46 $\pm$ 0.06 & 898.3 $\pm$ 1.8 \\
    GRB220514B & 10 & 49.94 $\pm$ 0.42 & 0.04 $\pm$ 0.02 & 5.50 $\pm$ 1.72 & 2.75 $\pm$ 0.06 & 683.4 $\pm$ 0.1 
    \enddata
    \centering
\end{deluxetable*}

\vspace{-0.8cm}

\begin{deluxetable*}{cccccc}[t!]
    \centering
    \tabletypesize{\scriptsize}
    \tablecaption{Number of astrophysical contaminants detected in the fields of each candidate through 30 trials occurring at random at least 2~hours from the trigger. Each of these flares would pass our detection criteria discussed in Section~\ref{subsec:pipe}. \label{tab:Contam}}
    
    \tablehead{
    \colhead{GRB Name} & \colhead{Galactic Latitude} & \multicolumn{2}{c}{Search Area} & \colhead{Flares} & \colhead{Contamination Fraction}\\
    \cline{3-4}
    \colhead{} & \colhead{(\degree)} & \colhead{(px)} & \colhead{(arcmin$^2$)} & \colhead{} & \colhead{(\%)}
    } 
    
    \startdata
    GRB180807A & -63.58 & 11751270 & 188 & 0 & 0\\ 
    GRB181208A & -21.93 & 3055459  &  49 & 2 & 7 \\
    GRB190117B & -12.25 & 6114157  &  98 & 3 & 10\\ 
    GRB190308A &  37.42 & 10086179 & 161 & 2 & 7\\ 
    GRB190723A &  13.89 & 11616207 & 186 & 1 & 3\\ 
    GRB200111A &  13.35 & 5217896  &  83 & 0 & 0\\ 
    GRB200412B & 25.92 & 117925 & 3 & 0 & 0\\
    GRB210114A &  8.063 & 14504152 & 232 & 2 & 7\\
    GRB210317A & -7.259 & 1511268  &  24 & 1 & 3\\
    GRB220310A &  67.44 & 37632    &   1 & 0 & 0\\
    GRB220514B &  40.85 & 387338   &   6 & 0 & 0\\ 
    \enddata
    \centering
\end{deluxetable*}

\vspace{-0.8cm}

\subsection{Contamination fraction} \label{subsec:contamFrac}

Contaminants are always present in transient searches, manifesting from a myriad of sources, including instrumental artefacts, asteroids, and flare stars. We can clearly identify instrumental artefacts by checking for bad subtractions and dubious pixel clustering; asteroids can also be clearly identified by their on-sky motion. However, flare stars are more challenging to rule out as they can evolve on similar timescales to GRB afterglows and in some cases have a similar light curve. 

In this search, we limited ourselves to identifying signals that occur within 1~hour of the GRB trigger, and within the $2\sigma$ error region. To estimate the contamination fraction of each candidate GRB field, we carry out 30 trial searches using evenly spaced times which are at least 2 hours from the GRB trigger. In each trial we record all transients that pass our selection criteria and calculate the contamination fraction to be the total number of transients detected divided by the number of test searches. The contamination fraction for each GRB field is summarised in Table~\ref{tab:Contam}. 

\begin{figure}[t]
    \centering
    \includegraphics[width=0.45\textwidth]{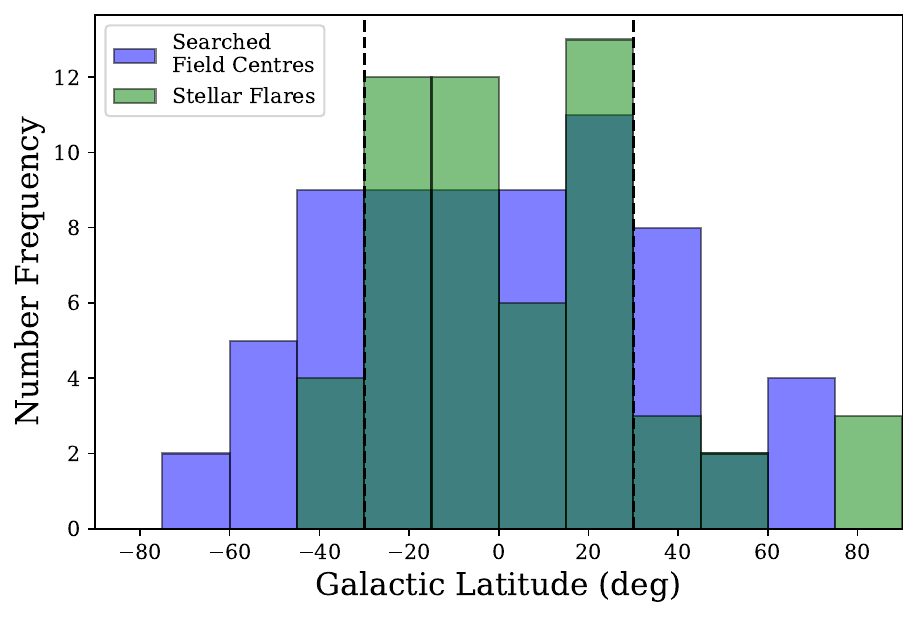}
    \caption{Distribution of detected stellar flares over galactic latitude in random trials across 69 field centres. Note that displayed in blue are the latitudes of the centres of fields; as shown in the far right bin, flares can be detected far from these centres.} 
    \label{fig:contamHist}
\end{figure}

For greater context detailing the relationship between contamination rate and galactic latitude, we repeat the test for each non-detected field outlined in Table~\ref{tab:limits} (see appendix). Of the 57 fields, 21 contained at least one contaminant, with a total number of 55 events. As shown by Fig.~\ref{fig:contamHist}, we find that the contamination fraction correlates with galactic latitude; this is expected due to the higher concentration of flare stars in the galactic plane. It should be noted that the apparent dip in flare detections around zero degrees latitude is not due to physical influences; denser regions of the sky have brighter limiting magnitudes in the reduction process, and thus flares must be brighter to overcome the background.

\vspace{-1.2cm}

\subsection{Non-Detections}

For GRB fields where we find no candidate, we estimate the limiting magnitude and coverage area of the 2$\sigma$ error region. These limits are presented in Table~\ref{tab:limits} (see appendix). As our condition for detection requires at least two consecutive frames $4\sigma$ brighter than the background, our limiting magnitude is dependent on the light curve shape and decay time. This means that the true limiting magnitude for a GRB afterglow is dependent on the observation cadence. To estimate the true limiting magnitude, we inject a template GRB afterglow light curve of our best-fit model for GRB220310A, which is our best-constrained GRB afterglow. We scale the flux until it is detected according to the \texttt{TESSreduce} zero point. For large error regions with multiple cutouts, we find that the zero point is variable across the field, and thus take the median value for this estimated limit. This issue is not applicable to small cutouts, which have a consistent zero point.

We combine all the pixels' limiting magnitudes into a single estimate through a weighted sum. We weight the contribution of each pixel according to a normalised 2D Gaussian centered on the reported burst position and using the 1$\sigma$ position errors as the standard deviation. This method incorporates the large positional uncertainty into our estimated magnitude limits. Due to field crowding and reduction quality, we find that the limit can vary significantly between a magnitude range of 13.5 and 18.4, where pixels containing stars have brighter limits than sky pixels.

\section{Discussion} \label{sec:discussion}

\subsection{Detection Confidence} \label{subsec:confidence}

From our sample of 11 new GRB afterglow candidates, we identify 3 as high-quality, 5 as tentative, and 3 as unlikely, as summarised in Table~\ref{tab:confidence}. This classification is based on four metrics: 

\begin{enumerate}
    \item We consider the amount of contamination in the candidate's field, as outlined in Section~\ref{subsec:contamFrac}. We give a passing grade (denoted by \checkmark) to fields with zero contamination, a partial grade (denoted by $\sim$) to those with contamination of up to 5\%, and a failed grade (denoted by \xmark) to fields with contamination above this.
    \item We inspect the full light curve of the candidate's coordinates to check for any other apparent outbursts across the entire \textit{TESS} sector. Light curves with no other visible signals earn a passing mark, while all other cases earn a failing mark. 
    \item We consider candidates with galactic latitude within $\pm 30\degree$ to be near/inside the galactic plane, and thus more likely to be flare stars as shown by Fig.~\ref{fig:contamHist}. These candidates are given a failing grade; all others a passing grade, other than GRB200412B, for which we give a partial grade as it lies within 5$\degree$ of the boundary. 
    \item We inspect the field crowding of the region surrounding the candidate's coordinates using PanSTARRS and SkyMapper imagery. Fields with no visible sources receive a passing grade and fields containing sources very near the coordinates receive a failing grade. Fields earn a partial grade if there are sources nearby that may or may not contaminate the pixels of the candidate; the large pixel size of \textit{TESS} makes this estimate uncertain.
\end{enumerate}

\begin{deluxetable*}{cccccc}[t!]
    \centering
    \tabletypesize{\scriptsize}
    \tablecaption{Candidate GRB afterglow confidence classification. Confidence is measured on performance in 4 categories: low contamination fraction (see Section~\ref{subsec:contamFrac}), solitary outburst occurrence across \textit{TESS} sector, galactic latitude falling outside Milky Way plane, and empty field in comparison with PanSTARRS and SkyMapper imaging.\label{tab:confidence}} 
    
    \tablehead{
    \colhead{GRB Name} & \multicolumn{4}{c}{Confidence Metric} & \colhead{Judgement}\\
    \cline{2-5}
    \colhead{} & \colhead{Low Contamination} & \colhead{Single Outburst} & \colhead{Galactic Latitude} & \colhead{Empty Visual Field} & \colhead{}
    } 
    \startdata
    GRB180807A & \checkmark & \checkmark & \checkmark & \checkmark & \checkmark\\ 
    GRB200412B & \checkmark & \checkmark & $\sim$ & \checkmark & \checkmark\\ 
    GRB220310A & \checkmark & \checkmark & \checkmark & \checkmark & \checkmark\\ 
    \cline{1-6}
    GRB181208A & \xmark & \checkmark & \xmark & $\sim$ & $\sim$\\
    GRB200111A & \checkmark & \checkmark & \xmark & \xmark & $\sim$\\
    GRB210114A & \xmark &\checkmark & \xmark & $\sim$ & $\sim$\\
    GRB210317A & $\sim$ & \checkmark & \xmark & $\sim$ & $\sim$\\
    GRB220514B & \checkmark & \xmark & \checkmark & $\sim$ & $\sim$\\
    GRB190117B & \xmark & \checkmark & \xmark & \xmark & \xmark\\
    GRB190308A & \xmark & \xmark & \checkmark & \xmark & \xmark\\
    GRB190723A & $\sim$ & \xmark & \xmark & \xmark & \xmark\\
    \enddata
    \centering
\end{deluxetable*}

\vspace{-0.5cm}

Our classifications can be further supported by other circulars on the GCN. Both GRB200412B and GRB220310A have had other circulars reporting detections of optical afterglows that agree with the locations we find \citep{Lip200412,Kum20,Bel20,Ste20,Xin20,Oga20,Neg22,Svi22,Lip22,Kum22,Fou22,Hos22}, whereas GRB180807A has no other circulars reported beyond the Fermi-GBM catalogue. GRB190723A, GRB200111A, and GRB210317A each were observed by multiple $\gamma$-ray telescopes with general consistency in location estimates to those reported by Fermi-GBM \citep{Lip19,Fer2020,Xia20,Koz20,Gai20,Pal20,Fer21,Yi21,Lip21}. One telescope observed an X-ray afterglow for GRB220514B \citep{Kaw22}, though no further positional information is reported to aid in localisation. The remaining candidates were only observed by the respective telescope that reported their bursts, and thus no comparison can be made beyond the coordinates utilised in this search.  

Upon reference to imaging catalogues, two candidates appear to be located near known sources. According to SIMBAD \citep{Wen00}, the presented coordinates for GRB190117B's candidate lie within 10~arcseconds of a potential flare star progenitor Gaia-DR3-2926429675703704448. Similarly, our GRB200111A candidate lies approximately 14~arcseconds away from the pulsating variable star ATO J099.2975+37.0822; this star's underlying variability appears to manifest in the long-term light curve of the pixels included within our candidate. The remaining candidates with partial or failed grades in the Empty Visual Field category of our assessment metric appear close to sources that are not named in any catalogue. We also investigate ATLAS photometry for candidate coordinates where possible, and find that no outbursts have been captured for the candidates of GRB190117B, GRB190308A, GRB200111A, and GRB220514B.

From this analysis, we conclude that the candidate afterglows for GRB180807A, GRB200412B, and GRB220310A originate from their respective GRBs. The candidate afterglows of GRB181208A, GRB200111A, GRB210114A, GRB210317A, and GRB220514B each exhibit at least one feature that inspires reasonable doubt in their legitimacy, and thus are tentative candidates. The candidate afterglows for GRB190117B, GRB190308A, and GRB190723A each exhibit a number of compelling features against their legitimacy, and thus are unlikely candidates.

\subsection{\textit{TESS} and GRB afterglows}

Our directed GRB afterglow survey shows promise for the capacity for \textit{TESS} to serendipitously detect rapid extragalactic transients. \cite{Smi21} predicted an observation rate of approximately 1 GRB per year; this was based on the general occurrence rate of GRBs, alongside the sky coverage and seeing ability of \textit{TESS}. Our study finds a rate consistent with this, though upper estimates reach $\sim2$ GRB afterglows per year when including all candidates presented in this paper. Additionally, we expect that moving forward, the rate of \textit{TESS}' detection of GRB afterglows will further rise due to its increased cadence of observation, allowing for the resolution of shorter signals. 

The real value of these observations, however, are their uniqueness. As highlighted above, only GRB200412B and GRB220310A have had their optical counterparts detected by other telescopes. As we are confident in our candidate for GRB180807A, we believe that this means \textit{TESS} may have been the only telescope to observe its afterglow 
It is likely that this is due to their poor localisation as reported by the detecting telescope, displaying the value of \textit{TESS}' wide FoV. Such an observation reveals the valuable role \textit{TESS} can play in the analysis of rapid transients with poor localisation.

\subsection{Modelling Optical GRB Afterglows}

The high-cadence observations conducted by the \textit{TESS} mission give a unique sample for modelling GRB afterglows. However, relying solely on the high-cadence observations taken with only the single \textit{TESS}-Red broadband filter is insufficient in extracting the modelling degeneracies in \texttt{afterglowpy}.

Despite the robustness of the five-parameter modelling procedure in producing well-constrained parameter estimates\footnote{These constraints are well-constrained for the five-parameter case, but one should be careful when extending these parameters to a more complex model.}, the inclusion of bootstrap sampling within these parameter spaces can result in light curves exhibiting significant residuals compared to the fitted model line\footnote{In this case the least-squares regression model was used to minimise the residual and compare by, $s = \sum (y_i - \hat{y}_i)^2$, where $y_i$ is the model data, and $\hat{y}_i$ is the real data.}. In order to obtain parameter-bound estimates during the modelling process, curves exhibiting a least-squares regression value higher than a predetermined threshold (which depends on the least-squares regression of the best fit) were excluded as potential candidates.

In particular, the key parameters $z_{\rm cmb}$ and $E_0$ exhibit a high level of degeneracy, necessitating observations at different wavelengths to disentangle them. In the event that the redshift had tighter constraints, for example, if the host galaxy was resolved, a more optimised set of parameters could be modelled. In an ideal scenario, simultaneous observations from a broadband blue variant of \textit{TESS} alongside the existing \textit{TESS} system would provide an extremely powerful data set for understanding GRBs and other transient-like phenomena.

Because of the non-linear nature of the model and the degeneracies present, the parameters we report are simply those that were found to have the lowest residual and parameter uncertainty; therefore, they are not to be taken as idealised constraints of the whole system. As some of the optimised parameters provide unrealistically small errors we implement a floor uncertainty of $0.01$ for $z_{\rm cmb}$, $0.1\,$s for $t_0$, as well as $0.001\,$radians for both $\theta_{\rm c}$ and $\theta_{\rm obs}$. 

Crucially, we found that an additional `afterglow onset delay' parameter, $t_0$, was required to shift the afterglow onset time from the prompt emission time (given by the time of trigger). This parameter accounts for \texttt{afterglowpy}'s lack of initial coasting phase - the time taken for the relativistic outflow to begin decelerating during its interaction with the CBM and subsequently release broadband synchrotron radiation. In our sample, we find a significant average onset delay of 740 $\pm$ 690 s; this is still present when considering only the high-quality afterglows of GRB180807A, GRB191016A, GRB200412B, and GRB220310A, giving an average delay of 560 $\pm$ 320 s. While we restrict our general modelling to a `top hat' geometry, the $t_0$ parameter was also required for the Gaussian jet geometry. 

The necessity of such a parameter in \texttt{afterglowpy} is clearly paramount for early afterglow light curve analysis. However, complications arise when trying to get a sense of the accuracy of the modelled $t_0$ values; the large spread in delay times may be indicative of different underlying causes between sources. A number of factors affect the observed onset of the afterglows. Many values fall within a single exposure of the reported detection time, including that of GRB210317A, which has the only `negative' onset. In these cases the light curve is simply too under-sampled to glean any conclusive result. Furthermore, as \texttt{afterglowpy} does not consider the sensitivity of the observing instrument, any input baseline flux is considered `zero flux', inherently linking the value of $t_0$ to the limiting magnitude of \textit{TESS}.

Despite this, the delays are not just sampling defects; GRB220310A provides a clear example of an afterglow non-detection at least 300 s post prompt emission, and a number of other candidate afterglows are also delayed beyond a single exposure. 

GRB afterglows are expected to arise some `short' time after the prompt emission due to the coasting phase of the relativistic outflow \citep{Kobayashi_2007}. As a consequence of both the historical lack of rapid follow up capability and the limitations outlined above, most studies consider the afterglow peak time, $t_p$, a more consistent probe for physical theories. Surveys documenting values for $t_p$ \citep[e.g.][]{Oates,Ghirlanda} have generally found a broad distribution with average measurements around 200~s. 

Each of our candidates appear to peak at times within the distributions reported by \cite{Ghirlanda}, though several lie close to the late end of the tail. For unlikely candidates such as GRB190308A and GRB190723A, such late peaks may provide further evidence against their legitimacy, though both GRB191016A and GRB220310A peak at late times as well. 

The explicit optical afterglow onset time itself - that is, the time between prompt emission and the very first optical signal - is a parameter which has not been empirically investigated in depth, simply due to observational limitations. Consequently, resources such as \texttt{afterglowpy} have not been required to incorporate onset delays in order to be viable tools for analysis; \textit{TESS} now challenges this paradigm.

\section{Conclusions}

We present the findings of the first systematic search for optical GRB afterglows in archival \textit{TESS} data. Our pipeline utilises \textit{TESS}' wide FoV to survey the fields of poorly-localised afterglows; upon analysis of 69 potential fields, we present 11 candidate signals alongside one previously documented afterglow. Of these 11, we have high confidence in the legitimacy of 3 and some level of uncertainty in the remainder. We attempt to model these light curves using the \texttt{afterglowpy} package for \texttt{python}, fitted through a range of sampling and least-squares regression techniques. We find that high-cadence \textit{TESS} broadband observations are not sufficient in breaking the degeneracy of key parameters, leading to poorly constrained fits and parameter estimates. However, we also find that there is a measurable delay time between the initial GRB burst and the afterglow onset that is not accounted for in \texttt{afterglowpy} - $740\pm690\,$s when considering all 12 events, and $560\pm320\,$s when only considering the four high-likelihood afterglows of GRB180807A, GRB191016A, GRB200412B, and GRB220310A. 

Only two of the eleven candidates have had other afterglow detections reported on the GCN, likely due to the localisation uncertainty released by the respective alert telescopes. \textit{TESS} therefore was likely the only optical telescope to observe the remaining events, including the probable afterglow of GRB180807A. This displays the capability \textit{TESS} has at filling a valuable role in the tranisent community for detecting poorly localised rapid transients as a direct result of its unique combination of near-continuous observation and a large FoV. Furthermore, with \textit{TESS} now operating at a cadence of 200 seconds, it will only improve at sampling the rise of optical afterglows; this will better constrain the apparent time delay between prompt emission and afterglow onset, a largely undocumented timescale.

\begin{acknowledgments}
This work was supported by NASA \textit{TESS} GI 80NSSC21K0242 (cycle 3). RRH was supported by the Marsden Fund Council from Government funding, managed by Royal Society Te Apārangi under the Fast Start Grant MFP-UOC2204.
This paper includes data collected by the \textit{TESS} mission. Funding for the \textit{TESS} mission is provided by the NASA’s Science Mission Directorate. \textit{TESS} data in this paper were obtained from the Mikulski Archive for Space Telescopes (MAST) at the Space Telescope Science Institute. This material is based upon work supported by NASA under award number 80GSFC21M0002.

The authors would like to acknowledge B. Gompertz and C. Kilpatrick for useful discussions. Q.W. is supported in part by NASA grants 80NSSC22K0494, 80NSSC21K0242 and 80NSSC19K0112. Q.W. is also partially supported by STScI DDRF fund.

\end{acknowledgments}

\bibliographystyle{aasjournal}
\bibliography{bibliography}

\onecolumngrid
\appendix

\section{Detection images}

\begin{figure}[H]
    \centering
    \plottwo{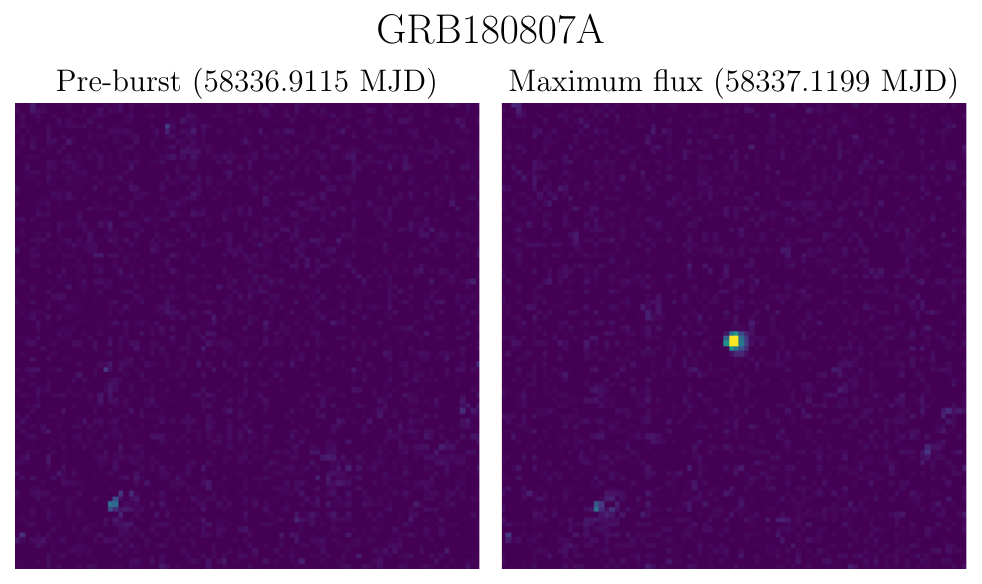}{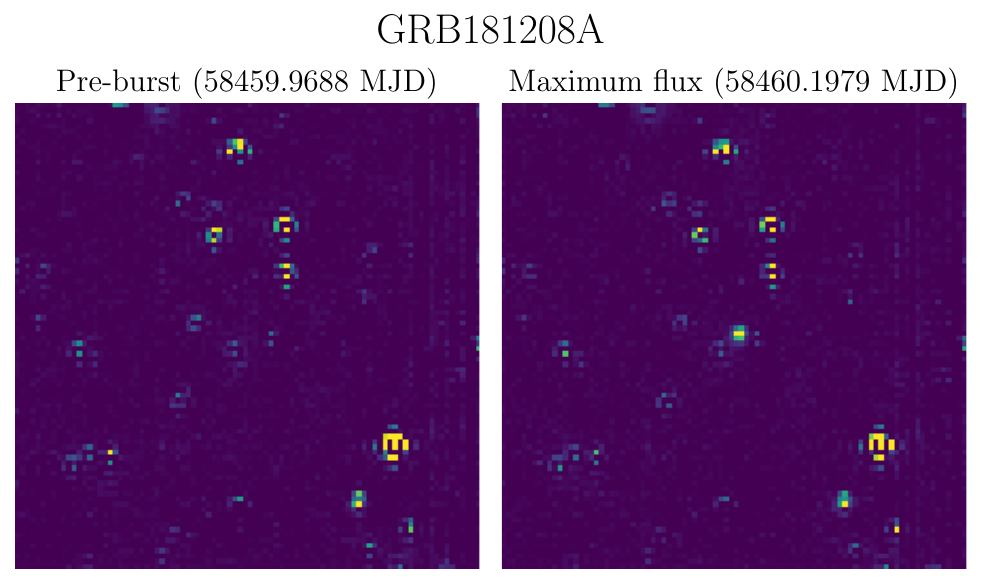}
    \plottwo{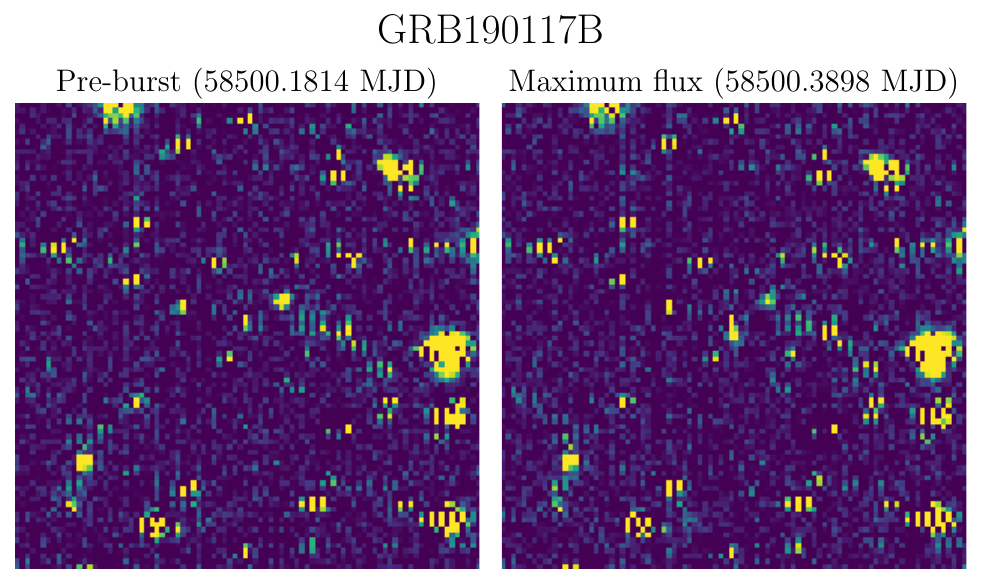}{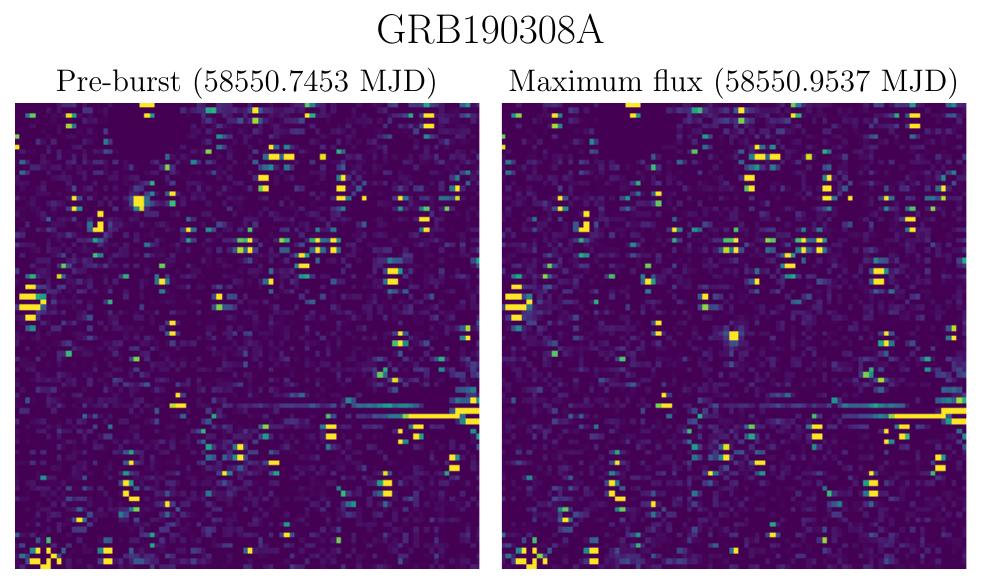}
    \plottwo{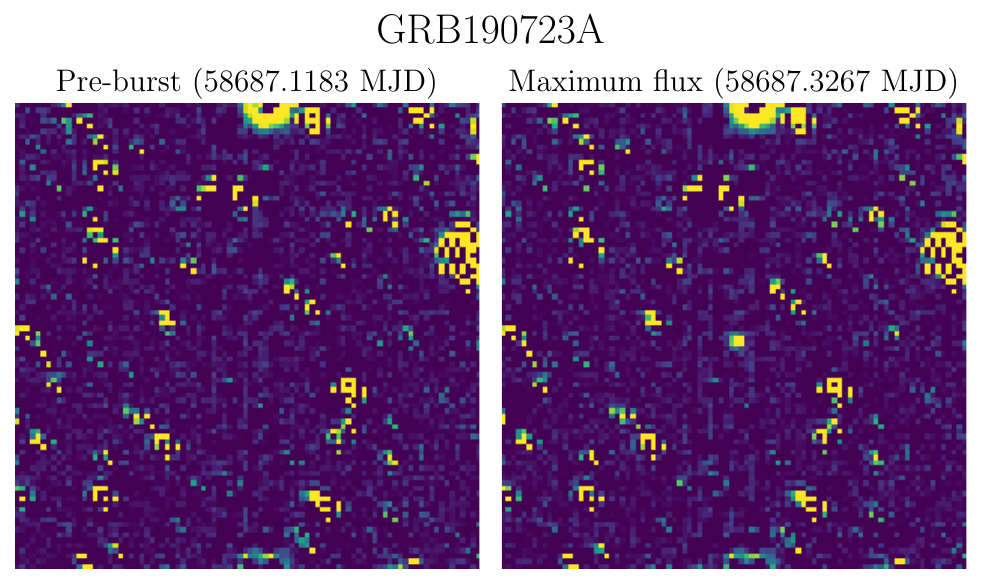}{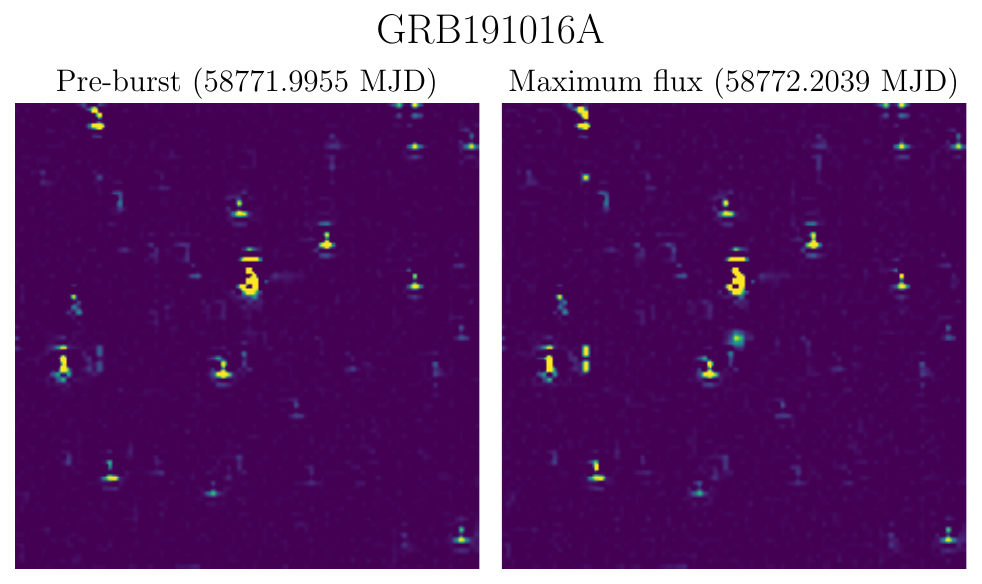}
    \plottwo{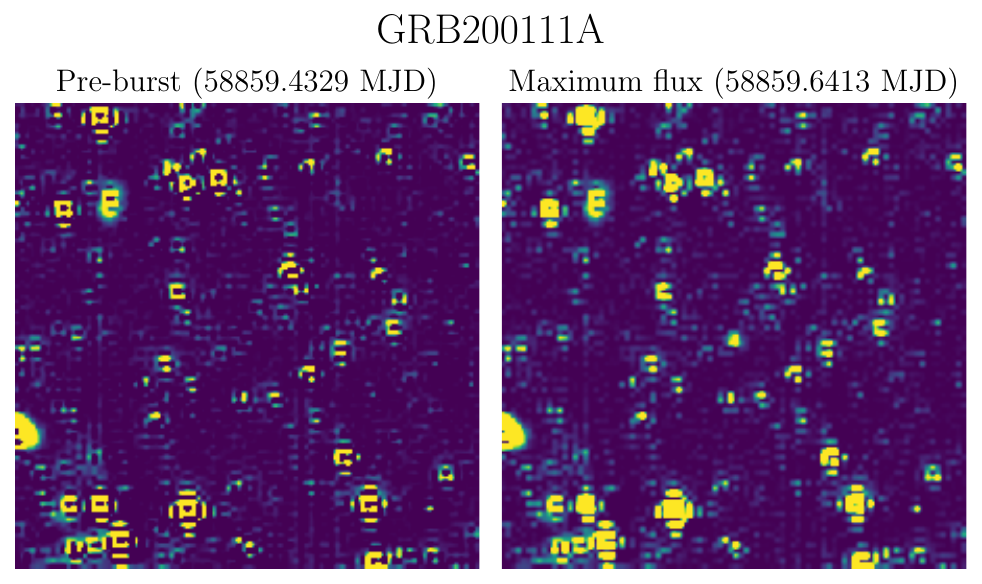}{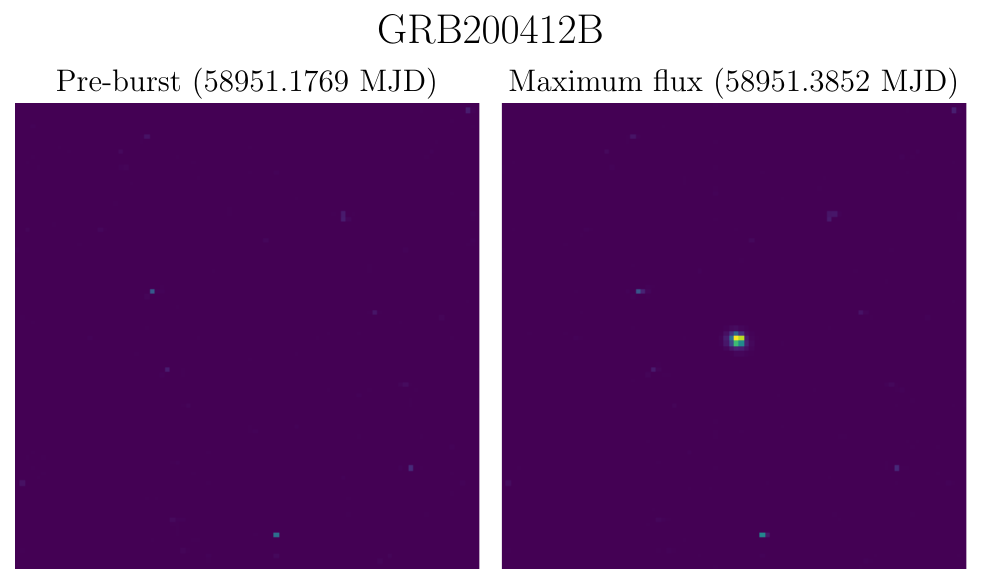}
    \caption{Detection images for all GRB afterglow candidates analysed in this study. For each GRB we show the difference images of the frame 10 cadences prior to burst trigger (left) and the frame with the highest flux (right).}
    \label{fig:max_flux_a}
\end{figure}

\begin{figure}[H]
    \centering
    \plottwo{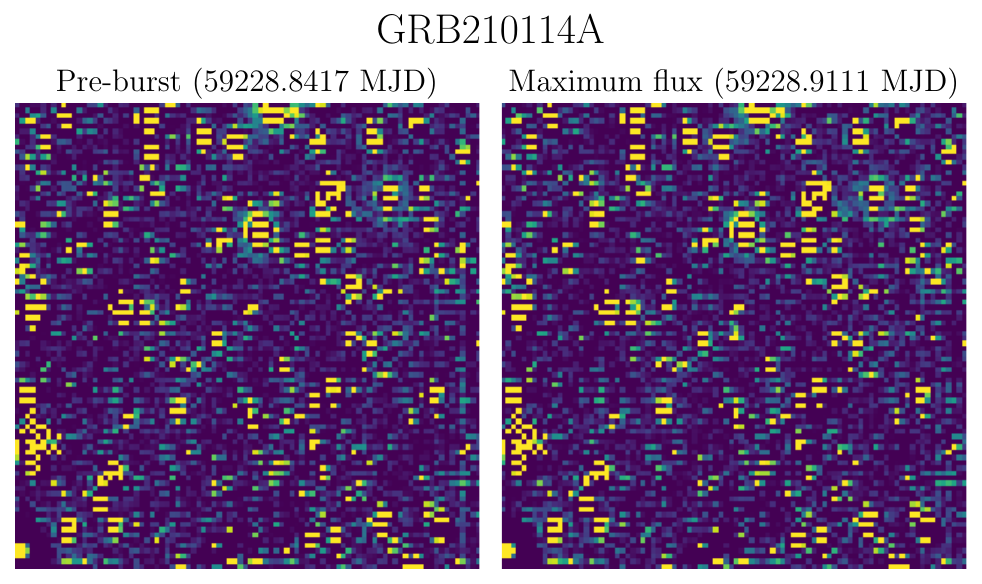}{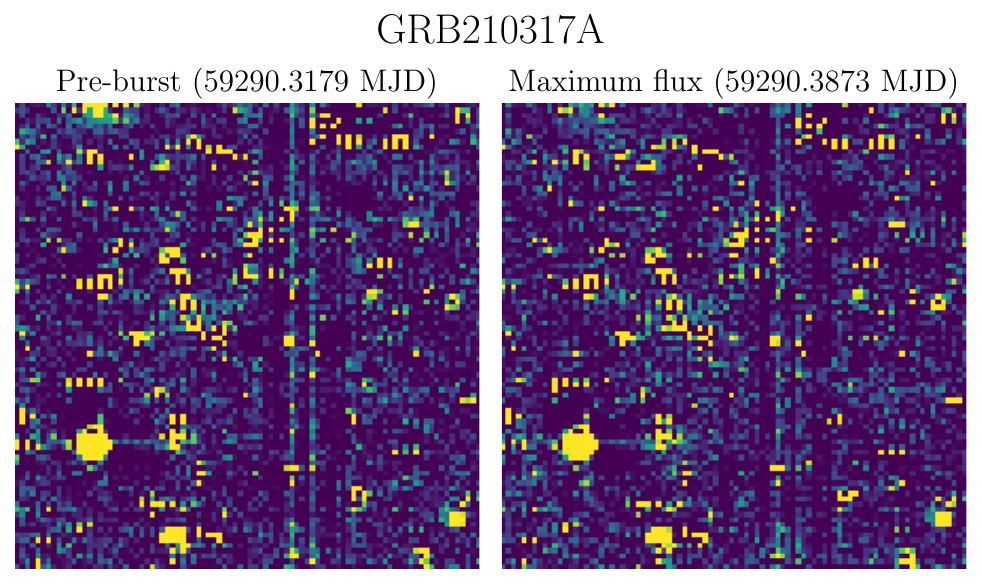}
    \plottwo{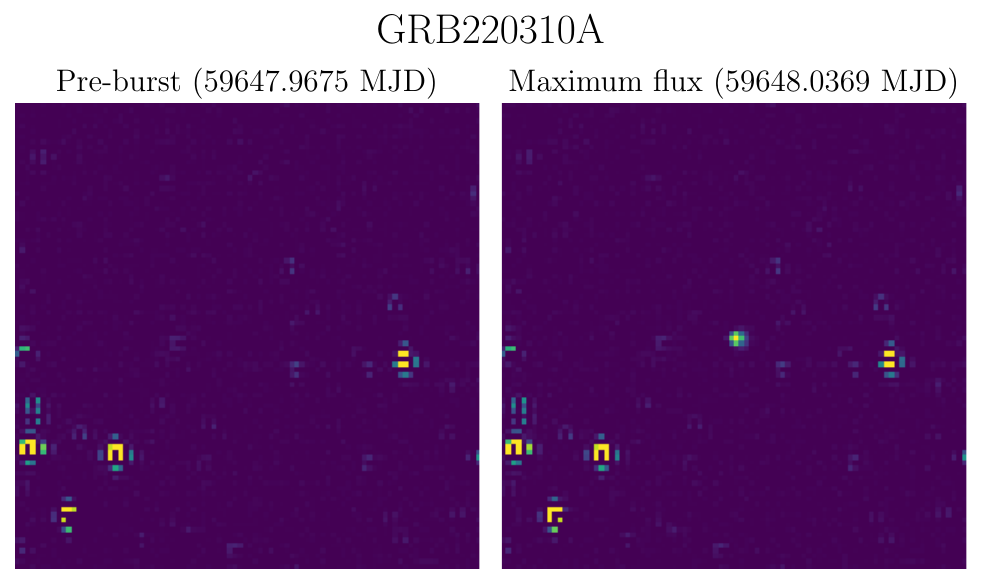}{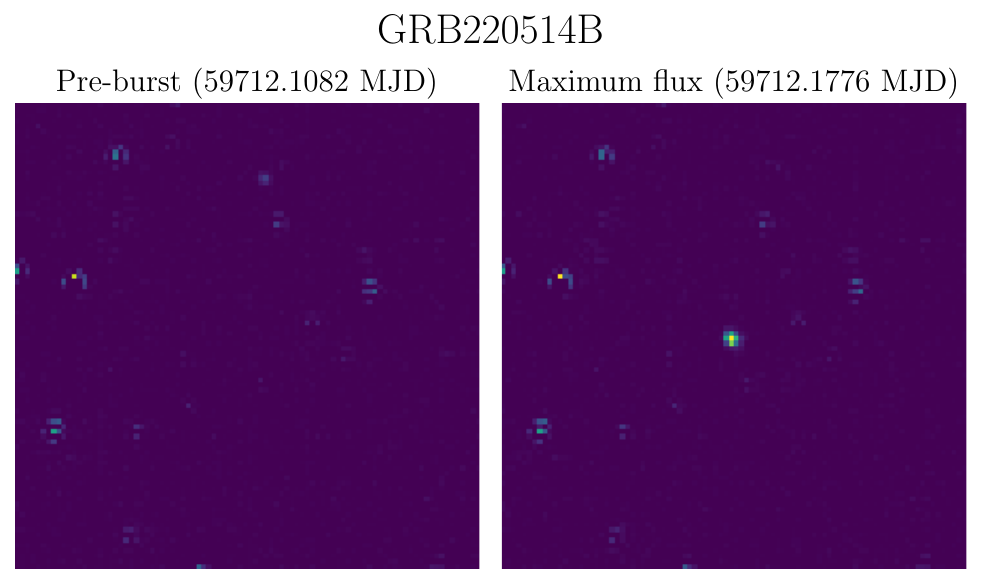}
    \par\medskip
    \textbf{Figure 6.} cont.
    \label{fig:max_flux_b}
\end{figure}

\newpage

\section{Non-detection limiting magnitudes} \label{sec:limiting_mag}
\begin{deluxetable}{cccccccccc}[h]
    \centering
    \tabletypesize{\notsotiny}
    \tablecaption{Average limiting magnitudes and coverage for coincident GRBs with non-detections in \textit{TESS}. The data is separated into three sections based on observation cadences of 30 m (ending with GRB200530A), 10 m (ending with GRB220823B), and 200 s respectively. \label{tab:limits}}
    
    \tablehead{
    \colhead{GRB Name} & \colhead{Detection Time} & \multicolumn{2}{l}{Estimated Position} & \colhead{Position Error} & \multicolumn{3}{c}{\textit{TESS} Coverage} & \colhead{2$\sigma$ Area Covered} & \colhead{Limiting Magnitude}\\
    \cline{3-4}
    \cline{6-8}
    \colhead{} & \colhead{(MJD)} & \multicolumn{1}{r}{R.A.} & \multicolumn{1}{c}{Dec.} & \colhead{($1\sigma$ , \degree)} 
    & \colhead{Sector} & \colhead{[Cam,CCD]} & \colhead{1st Obs - Trigger (s)}}
    \startdata
    GRB180727A & 58326.594074 & 346.67 & -63.05 & $\sim$ 0 & 1 & [2,1] & 361.1 & - & 15.1 \\ 
    GRB180801B & 58331.492149 & 3.38 & -70.68 & 3.877 & 1 & [3,2],[3,1] & 85.1 & 92.9\% & 16.4 \\ 
    GRB180804D & 58334.930960 & 109.36 & -69.32 & 3.506 & 1 & [4,3],[4,2] & 1464.3 & 97.6\% & 16.1 \\ 
    GRB180821B & 58351.653208 & 309.78 & -38.12 & 4.966 & 1 & [1,1],[1,4] & 247.2 & 37.9\% & 17.3 \\
    GRB180906A$^\dag$ & 58367.597127 & 104.81 & -67.02 & 3.619 & - & - & - & - & - \\ 
    GRB180922A & 58383.461013 & 19.18 & -0.13 & 4.462 & 3 & [1,3],[1,4] & 1184.2 & 44.9\% & 15.6 \\ 
    GRB180924A & 58385.640532 & 49.20 & -58.53 & $\sim$ 0 & 3 & [3,3] & 78.7 & - & 16.1 \\ 
    GRB181022A & 58413.729317 & 56.79 & -29.38 & $\sim$ 0 & 4 & [2,2] & 1253.1 & - & 15.9 \\ 
    GRB190102A & 58485.257993 & 86.39 & 16.69 & 4.021 & 6 & [1,4],[1,3] & 723.1 & 44.3\% & 16.0 \\ 
    GRB190117A & 58500.608877 & 113.86 & 6.52 & 3.367 & 7 & [1,4],[1,3],[1,1] & 1021.0 & 93.5\% & 16.6 \\ 
    GRB190129B & 58512.510810 & 117.28 & 1.25 & 2.033 & 7 & [1,1],[1,2],[1,4],[1,3] & 477.3 & 86.0\% & 16.1 \\ 
    GRB190407A & 58580.575420 & 90.53 & -64.14 & 3.894 & 10 & [4,3],[4,4],[4,2],[4,1] & 1573.0 & 90.8\% & 16.1 \\ 
    GRB190422C$^\dag$ & 58595.670061 & 189.23 & -54.95 & 6.281 & - & - & - & - & - \\ 
    GRB190507C & 58610.712000 & 86.22 & -61.88 & 13.394 & 11 & [4,2],[4,1],[4,3],[4,4] & 568.5 & 51.7\% & 13.4 \\ 
    GRB190630C & 58664.995127 & 293.88 & -32.74 & $\sim$ 0 & 13 & [1,3] & 45.5 & - & 14.6 \\ 
    GRB190726A & 58690.642269 & 310.26 & 34.29 & 3.843 & 14 & [1,1],[1,4] & 1519.1 & 51.8\% & 15.9 \\ 
    GRB191028B & 58784.588863 & 276.88 & 69.99 & 5.569 & 17 & [4,2],[4,1],[4,3],[4,4] & 517.2 & 92.9\% & 16.7 \\ 
    GRB191117A & 58804.005873 & 297.88 & 76.38 & 6.768 & 18 & [3,3],[3,4],[4,2],[4,1] & 577.3 & 92.7\% & 15.8 \\ 
    GRB200110A & 58858.518117 & 96.15 & 28.88 & 4.724 & 20 & [1,3] & 1498.6 & 23.2\% & 16.6 \\
    GRB200112A & 58860.525359 & 150.13 & 64.41 & 2.831 & 20 & [2,1] & 872.8 & 45.1\% & 17.7 \\
    GRB200303A & 58911.107211 & 212.72 & 51.36 & $\sim$ 0 & 22 & [3,2] & 914.7 & - & 15.9 \\ 
    GRB200324A & 58932.693831 & 222.67 & 35.94 & $\sim$ 0 & 23 & [2,1] & 702.8 & - & 15.7 \\ 
    GRB200403B & 58942.919392 & 239.90 & 73.24 & 3.370 & 23 & [4,1] & 1026.6 & 75.2\% & 17.1 \\ 
    GRB200530A & 58999.030671 & 32.10 & 69.23 & 0.483 & 25 & [4,3] & 1574.0 & 100\% & 15.8 \\ \hline
    GRB200901A & 59093.157957 & 61.78 & -59.89 & $\sim$ 0 & 29 & [3,4] & 18.0 & - & 16.4 \\ 
    GRB200909A$^\dag$ & 59101.167580 & 338.86 & -16.26 & 4.410 & - & - & -  & - & - \\
    GRB201104A$^\dag$ & 59157.000648 & 81.40 & -71.10 & 0.233 & - & - & - & - & - \\ 
    GRB201120B$^\dag$ & 59173.441533 & 78.16 & -57.70 & 13.110 & - & - & - & - & - \\ 
    GRB210126A$^\dag$ & 59240.416717 & 70.68 & -67.94 & 5.770 & - & - & - & - & - \\ 
    GRB210207A & 59252.602755 & 137.97 & 4.73 & 4.559 & 34 & [1,3],[1,2] & 322.7 & 38.2\% & 18.2 \\ 
    GRB210222B$^\dag$ & 59267.942662 & 154.61 & -14.93 & $\sim$ 0 & - & -  & -& - & - \\ 
    GRB210315A & 59288.715863 & 98.77 & -54.76 & 6.552 & 36 & [4,2],[4,1],[3,3],[4,3] & 323.4 & 43.3\% & 17.3 \\ 
    GRB210317A & 59290.380891 & 154.22 & -64.94 & 2.784 & 36 & [3,4],[3,1] & 463.3 & 63.6\% & 16.9 \\ 
    GRB210411B & 59315.564248 & 115.68 & -74.84 & 4.333 & 37 & [4,1],[4,2],[3,4] & 213.3 & 83.4\% & 17.4 \\ 
    GRB210419A & 59323.287280 & 86.85 & -65.50 & $\sim$ 0 & 37 & [4,3] & 150.5 & - & 15.5 \\ 
    GRB210427A & 59331.206389 & 177.79 & -52.87 & 3.600 & 37 & [2,2],[2,1] & 545.3 & 90.8\% & 16.5 \\ 
    GRB210504A & 59338.579780 & 222.39 & -30.53 & $\sim$ 0 & 38 & [1,4] & 511.9 & - & 15.9 \\ 
    GRB210520A & 59354.796551 & 123.00 & -69.40 & 0.733 & 38 & [4,2] & 369.1 & 91.2\% & 17.3 \\
    GRB210621A & 59386.447257 & 248.31 & -61.07 & 5.059 & 39 & [2,4],[2,3],[1,1],[2,1],[2,2] & 459.6 & 90.7\% & 16.2 \\ 
    GRB210730A & 59425.206586 & 149.59 & 69.69 & $\sim$ 0 & 41 & [4,3] & 153.7 & - & 16.0 \\ 
    GRB210928B$^\dag$ & 59485.761173 & 43.86 & 22.66 & 5.969 & - & - & - &  - & - \\ 
    GRB211225A$^\dag$ & 59573.014202 & 124.67 & 16.81 & 6.025 & - & - & -  & - & - \\ 
    GRB220319A & 59657.736493 & 218.22 & 61.30 & $\sim$ 0 & 49 & [3,4] & 410.7 & - & 14.9 \\ 
    GRB220321A & 59659.282187 & 170.76 & 28.94 & 7.707 & 49 & [1,2],[1,1],[1,3] & 61.0 & 36.8\% & 18.4 \\ 
    GRB220424A & 59693.481125 & 235.23 & 47.86 & 4.703 & 51 & [3,1],[3,4] & 388.0 & 92.3\% & 17.8 \\ 
    GRB220617B & 59747.672847 & 284.49 & 25.59 & 7.342 & 53 & [1,1],[1,2],[2,4],[1,4],[1,3] & 70.2 & 43.8\% & 16.5 \\
    GRB220623A & 59753.294532 & 145.40 & 75.82 & $\sim$ 0 & 53 & [4,4] & 348.8 & - & 16.3 \\ 
    GRB220708A & 59768.194086 & 103.13 & 72.14 & $\sim$ 0 & 53 & [4,4] & 18.3 & - & 16.4 \\ 
    GRB220810A & 59801.969222 & 309.35 & 2.66 & 2.808 & 55 & [1,2],[1,3] & 600.0 & 40.66\% & 17.1 \\
    GRB220823B & 59814.618232 & 287.96 & 63.73 & 6.648 & 55 & [4,2],[4,1],[4,3],[3,3],[3,4] & 336.9 & 86.0\% & 17.1 \\ \hline
    GRB220924A & 59846.906123 & 321.48 & 33.84 & 4.617 & 56 & [2,2],[2,3] & 110.1 & 67.1\% & 14.3 \\ 
    GRB221004A & 59856.622349 & 356.83 & 57.71 & 7.938 & 57 & [2,1],[2,2],[2,3],[2,4],[3,2] & 27.3 & 75.7\% & 14.4 \\ 
    GRB221029A$^\dag$ & 59881.045455 & 53.25 & 44.88 & 4.455 & - & - & - & - & - \\ 
    GRB221120A & 59903.895460 & 41.34 & 43.24 & $\sim$ 0 & 58 & [1,1] & 164.9 & - & 13.5 \\ 
    GRB221221A & 59934.944097 & 81.82 & 63.40 & 3.280 & 59 & [2,4],[2,1] & 111.1 & 98.0\% & 14.9 \\ 
    GRB230116D & 59960.878275 & 98.62 & 49.87 & $\sim$ 0 & 60 & [1,2] & 32.6 & - & 13.3 \\ 
    GRB230116E & 59960.575175 & 303.60 & 76.00 & 2.747 & 60 & [4,1],[3,4] & 20.8 & 94.6\% & 15.4 \\  
    \enddata
    \centering
    \tablenotetext{\dag}{Occurred during a \textit{TESS} downlink.}
\end{deluxetable}
\vspace{-100cm}
\end{document}